\documentclass[]{article}
\usepackage[utf8]{inputenc}
\usepackage{amsthm}
\usepackage{amsmath,amssymb}
\usepackage[square,numbers,sort&compress]{natbib}
\usepackage[strings]{underscore}
\usepackage{lmodern}
\usepackage{listings}

\usepackage[a4paper,
            bindingoffset=0.2in,
            left=1in,
            right=1in,
            top=1in,
            bottom=1in,
            footskip=.25in]{geometry}

\renewcommand{\baselinestretch}{1.5} 
\usepackage{mathtools} 

\usepackage{authblk}
\usepackage{appendix}

\newtheorem{theorem*}{Theorem}

\usepackage{mathtools}
\usepackage{mdframed}

\newmdtheoremenv{theo}{Theorem}

\usepackage{bbm}
\usepackage{mdframed}
\usepackage{xcolor,sectsty}

\definecolor{astral}{RGB}{46,116,181}

\let\oldstar\*
\renewcommand{\*}[1]{\mathbf{#1}}

\newcommand{\pp}[2]{\frac{\partial #1}{\partial #2}}

\newcommand{\fR}{\mathbb{R}}

\usepackage{color, colortbl}
\definecolor{Gray}{gray}{0.8}
\definecolor{LightCyan}{rgb}{0.88,1,1}
\definecolor{green(html/cssgreen)}{rgb}{0.0, 0.5, 0.0}

\newcommand{\ca}{$\text{Ca}^{2+}\ $}

\newmdenv[
  backgroundcolor=gray!10,  
  leftmargin=0pt,
  innerleftmargin=10pt,
  innerrightmargin=10pt,
  innertopmargin=10pt,
  innerbottommargin=10pt,
  linewidth=1pt,            
  linecolor=black,          
  topline=true,             
  bottomline=true,          
  rightline=true,           
  leftline=true             
]{mybox}

\let\oldstar\*
\renewcommand{\*}[1]{\mathbf{#1}}

\title{Neuron-Astrocyte Associative Memory}
\author[1]{Leo Kozachkov}
\author[1,2]{Jean-Jacques Slotine}
\author[3]{Dmitry Krotov}
\affil[1]{Department of Brain and Cognitive Sciences, MIT}
\affil[2]{Department of Mechanical Engineering, MIT}
\affil[3]{MIT-IBM Watson AI Lab, IBM Research}
\affil[ ]{\texttt{\{leokoz8,jjs\}@mit.edu, krotov@ibm.com}}
\date{}

\begin{document} 


\baselineskip24pt


\maketitle


\begin{abstract}
 Astrocytes, the most abundant type of glial cell, play a fundamental role in memory. Despite most hippocampal synapses being contacted by an astrocyte, there are no current theories that explain how neurons, synapses, and astrocytes might collectively contribute to memory function. We demonstrate that fundamental aspects of astrocyte morphology and physiology naturally lead to a dynamic, high-capacity associative memory system. The neuron-astrocyte networks generated by our framework are closely related to popular machine learning architectures known as Dense Associative Memories or Modern Hopfield Networks. In their known biological implementations the ratio of stored memories to the number of neurons remains constant, despite the growth of the network size. Our work demonstrates that neuron-astrocyte networks follow superior, supralinear memory scaling laws, outperforming all known biological implementations of Dense Associative Memory. This theoretical link suggests the exciting and previously unnoticed possibility that memories could be stored, at least in part, within astrocytes rather than solely in the synaptic weights between neurons. 
\end{abstract}


Not all brain cells are neurons. It is estimated that about half of the cells in the human brain are glial cells (from ``glue" in Greek) \cite{kandel2000principles}. Glial cells have long been known to play an important role in homeostatic brain functions, such as regulating blood flow \cite{witthoftBidirectionalModelCommunication2012} -- thus contributing to hemodynamic signals such as those measured in fMRI \cite{schummersTunedResponsesAstrocytes2008} -- and removing synaptic debris. Converging lines of recent evidence strongly suggest that they are also directly involved in learning, memory, and cognition \cite{mu2019glia,kastanenkaRoadmapIntegrateAstrocytes2020,kol2020astrocytes,agid2020glial,nagai2021behaviorally,requie2022astrocytes,rupprecht2024centripetal}. Among glial cells, \textit{astrocytes} are particularly important for brain function. They serve a crucial role in directly sensing neural activity and, in turn, regulating synaptic strength and plasticity \cite{kastanenkaRoadmapIntegrateAstrocytes2020,srinivasan2015ca2,mu2019glia,semyanov2021astrocytic,noh2023cortical,de2023specialized}. In addition to sensing neural activity, astrocytes are also important targets of neuromodulatory signals such as norepinephrine and acetylcholine emerging from potentially distant brain structures such as the brain stem \cite{murphy2023conceptual}. 

Of particular relevance to the computational neuroscience community are the recent findings that 1) astrocytes are necessary for forming long-term memories \cite{suzuki2011astrocyte,pinto2019impairments,kol2020astrocytes,sun2024spatial} and 2) astrocytes respond to neural activity on timescales spanning many orders of magnitude, from several hundred milliseconds to minutes \cite{di2011local,stobartCorticalCircuitActivity2018,de2023specialized}. Despite extensive evidence establishing the importance of neuron-astrocyte interactions for long-term memory function, computational theories of these interactions are still in their infancy.
\vspace{-0.2cm}
\paragraph{What Shapes Astrocytic Computation?}
The core proposal of this paper is that astrocytes compute, and these computations are shaped by tunable signalling pathways within astrocytes. We will be primarily concerned with associative computations: how neurons, synapses, and astrocytes work together to store and retrive memories. In this case, astrocytic \ca flux coefficients are the site of memory storage, and neuron-synapse-astrocyte interactions are the mechanism of memory retrieval. This proposal harmoniously extends decades of prior work suggesting that memories are stored in synaptic weights \cite{barrionuevo1983associative, markram1997regulation} and provides a new perspective where synaptic weights ``emerge" from interactions between neurons and astrocytes.  
\begin{figure}[ht!]
  \centering
  \includegraphics[width=\textwidth]{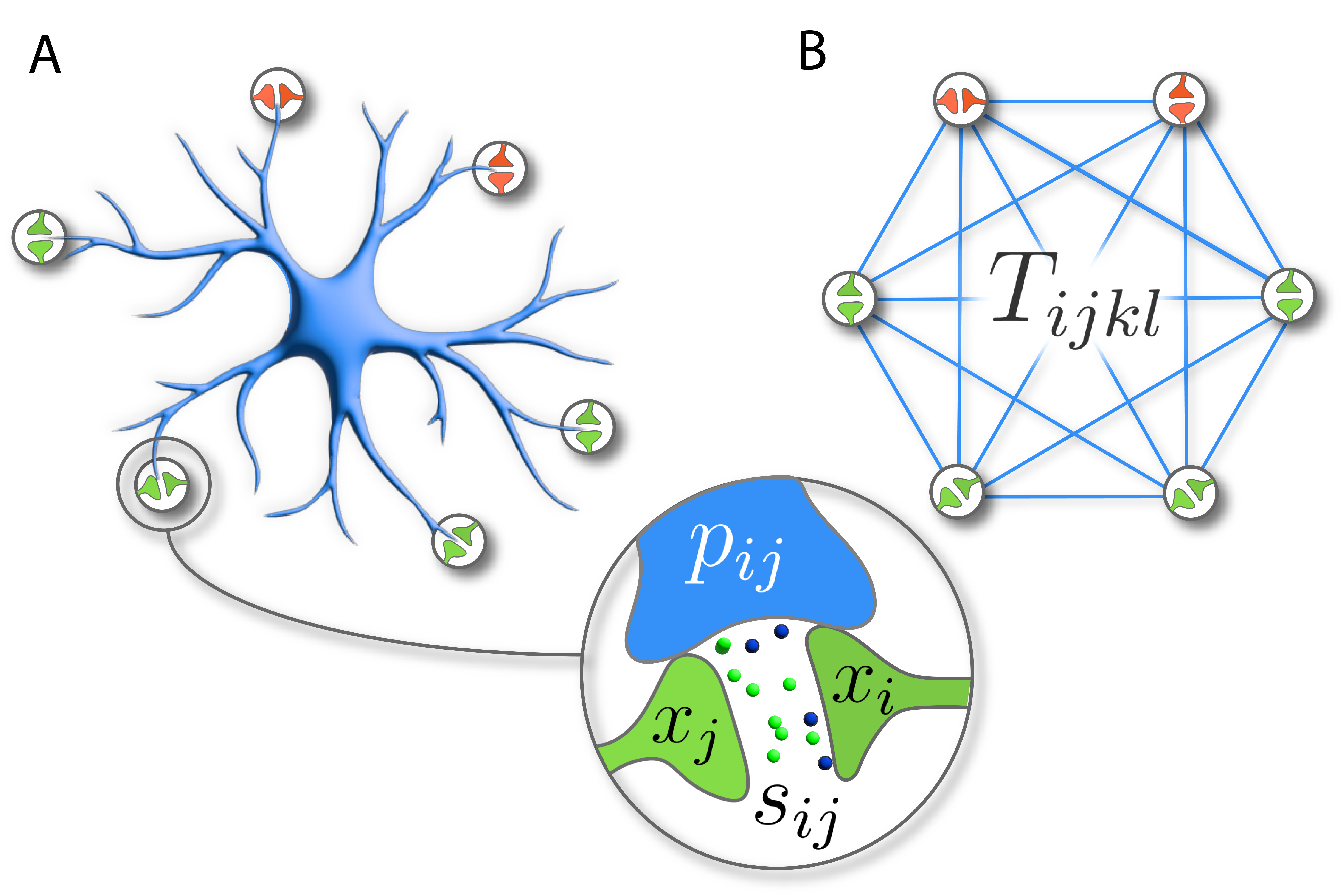}
  \caption{A) An abstracted version of an astrocyte, showing the astrocyte processes and the synapses. B) Our mathematical idealization of the mini-circuit defined by a single astrocyte.}
  \label{fig:cartoon}
\end{figure}
\section{Neuron-Astrocyte Model}\label{section:neuron-astrocyte-model}
\vspace{-0.2cm}
Astrocytes have a primary cell body (soma) with numerous branching processes that envelope nearby synapses (Figure \ref{fig:cartoon}). This three-part structure is known as the \textit{tripartite synapse} \cite{pereaTripartiteSynapsesAstrocytes2009}. A single astrocyte can form over $10^6$ tripartite synapses \cite{allen2017cell}, and astrocyte networks spatially tile the brain, forming non-overlapping ``islands" \cite{halassaSynapticIslandsDefined2007}. Astrocyte processes detect neurotransmitters in the synaptic cleft, leading to an upsurge in intracellular free calcium \ca ions within the astrocyte process. This leads to a biochemical cascade in the astrocyte, potentially culminating in the release of gliotransmitters back into the synaptic cleft, influencing neural activity--a closed feedback loop. Astrocyte processes can intercommunicate through calcium transport \cite{araqueGliotransmittersTravelTime2014}, and individual astrocytes connect via gap junctions. The interplay between neurons and astrocytes, spanning multiple temporal and spatial scales, underscores the relevance of astrocytes in learning and memory. For this paper, we will focus on the following salient aspects of astrocyte biology:
\begin{itemize}
\item A single astrocyte can connect to millions of nearby synapses, forming three-part connections (astrocyte process, pre-synaptic neuron, post-synaptic neuron) called tripartite synapses\cite{pereaTripartiteSynapsesAstrocytes2009}.
\item Astrocytes detect neural activity and respond by regulating this activity through the release of gliotransmitters \cite{de2016astrocytes}.
\item Tripartite synapses can interact with each other, possibly through astrocytic intracellular calcium transport \cite{araqueGliotransmittersTravelTime2014}.
\end{itemize}

\paragraph{Neural Dynamics}
The above points may be formalized into a set of dynamical equations governing the time evolution of neurons, synapses, and astrocytes. The membrane voltage $x_i$ for each neuron $i$ evolves according to a standard rate recurrent neural network model \cite{wilsonExcitatoryInhibitoryInteractions1972,hopfield1984neurons} with the characteristic time scale of the neural dynamics $\tau_n$, and the leak rate $\lambda$

\begin{equation}\label{eq:neuron_eqs}
\tau_n \ \dot{x}_i = -\lambda \ x_i \ + \ \sum^N_{j = 1} \ g(s_{ij})
\ \phi(x_j)  + b_i   
\end{equation}
Each neuron has an input $b_i$, which establishes the neuron's baseline activation. The nonlinearity $\phi(x_j)$ transforms neural membrane voltages into firing rates, while the term $g(s_{ij})$ indicates the strength of the synaptic weight connecting neurons $i$ and $j$. The variable $s_{ij}$ is dynamic and alters depending on the activity of both neurons and astrocytes, as we will detail next. Note that for fixed $s_{ij}$, this model is simply a standard recurrent neural network. 
\begin{figure}
\begin{mybox}
\textbf{Examples of Possible Lagrangians and Activations}
\begin{equation}
\begin{split}
&\mathcal{L}(\mathbf{z}) = \log \sum_{i = 1}^N e^{z_i} \quad \rightarrow \quad \pp{\mathcal{L}(\mathbf{z})}{\mathbf{z}} =  \text{Softmax}(\mathbf{z}) \\
&\mathcal{L}(\mathbf{z}) = \sum_{i = 1}^N Q(z_i) \quad \ 
  \rightarrow \quad \pp{\mathcal{L}(\mathbf{z})}{\mathbf{z}} =  \big[q(z_1), \hdots, q(z_n)\big]^T
\end{split}
\end{equation}
\end{mybox}
\caption{Examples of possible Lagrangian functions. Here the variable $\mathbf{z}$ is an arbitrary dynamical variable in our model (e.g., astrocyte calcium level). Recall from the main text that activation functions are defined from the Lagrangians as $\pp{\mathcal{L}}{z_i}$. The first Lagrangian provides an example of a ``collective" activation functions. The second Lagrangian leads to an element-wise activation function, assuming $\pp{Q}{z_i} = q(z_i)$. Generally, the only mathematical requirement for our Lagrangians is that they must be convex functions.}
\label{figure:lagrangians}
\end{figure}

\vspace{-0.2cm}
\paragraph{Synaptic Dynamics}
The level of synaptic facilitation, which is denoted by $s_{ij}$, refers to the degree to which pre-synaptic spiking activity impacts the post-synaptic neuron. Biophysically, there are many factors which influence the efficacy of a synapse, such as the number of postsynaptic receptors and the level of calcium in the synaptic cleft. The strength of the synapse can either increase or decrease based on both pre- and post-synaptic activity, as observed in Hebbian plasticity. As in earlier studies of neuron-glial interactions, we consider tripartite synapses--synapses whose plasticity is modulated by an enveloping astrocytic process, $p_{ij}$
\begin{equation}\label{eq:synaptic_dynamics}
\tau_s \ \dot{s}_{ij} = -\alpha \ s_{ij} + f(x_i,x_j, p_{ij}) + c_{ij}
\end{equation}
The timescale of the synaptic dynamics is $\tau_s$, and $\alpha$ establishes the leak-rate of synaptic facilitation. The function $f$ encapsulates the interactions between these three biological variables. The inputs $c_{ij}$ serve as bias variables, controlling the baseline rate of synaptic facilitation in the absence of external input. The concentration of intracellular \ca ions in the astrocytic process that wraps around the particular synapse $i-j$ is denoted by $p_{ij}$. Biophysically, astrocytes influence neural activity through \ca-dependent exocytosis of gliotransmitters such as GABA, D-serine, ATP and glutamate \cite{de2023specialized}--an influence which is encoded in the function $f$. The level of intracellular astroctytic \ca is a dynamic variable and its value depends on both the calcium levels in adjacent astrocyte processes and the synaptic state $s_{ij}$.
\vspace{-0.2cm}
\paragraph{Astrocyte Process Dynamics}
The state of a specific astrocytic process is determined by its interactions with neurons at the tripartite synapse, plus its interactions with other processes through intracellular calcium transport
\begin{equation}\label{eq:astro_eqs}
\tau_p \ \dot{p}_{ij} = -\gamma \ p_{ij} \ + \ \sum^N_{k,l = 1 } \ T_{ijkl} \ \psi(p_{kl})+ \kappa(s_{ij}) + d_{ij}
\end{equation}
The double sum in the astrocyte equations captures the interactions between process $p_{ij}$ and all other processes \cite{arizono2020structural}. In the simplest scenario, calcium can diffuse between processes, resulting in a linear function $\psi$ and tensor $T_{ijkl}$ describing concentration fluxes. More complex calcium transport mechanisms between different processes within an astrocyte can produce non-linear functions $\psi$. The term $d_{ij}$ is a constant bias term which sets the overall ``tone" of the astrocyte. This variable may be thought of as a neuromodulatory signal, potentially arriving from distant brain regions such as the pons.  The input from process $p_{kl}$ to process $p_{ij}$ is weighed by the scalar $T_{ijkl}$. A zero value for this scalar indicates no direct physical connection between process $ij$ and $kl$. Hence, the astrocyte's anatomical structure can be encoded in the non-zero entries of tensor $T$. The nonlinear function $\kappa$ encapsulates the synapse → astrocyte signalling pathway at the tripartite synapse level. Here, $\tau_p$ represents the astrocyte timescale, while $\gamma >0$ is a leak term for the intracellular calcium in the astrocyte process.



\vspace{-0.3cm}
\section{Associative Neuron-Astrocyte Model}
In section \ref{section:neuron-astrocyte-model}, we described a general framework, grounded in the biology of neuron-astrocyte communication, for modelling neuron-astrocyte interactions via the tripartite synapses. Depending on the choices of the nonlinearities and parameters this network can exhibit many sophisticated dynamical behaviours--such as chaos or limit cycles--which can be difficult to analyse in full generality. 

To better understand the potential role of neuron-astrocyte interactions, we will focus on an important limiting case where the system demonstrates associative memory functions. As in essentially all models of biological associative memory, this requires the presence of symmetries in the governining equations of the biological circuit. We show that under certain conditions the resulting neuron-astrocyte model has a global energy function (Lyapunov function), which monotonically decreases on the dynamical trajectory and is bounded from below. This makes it possible to identify a regime of operation of the neuron-astrocyte network that results in dynamical trajectories converging to fixed point attractor states. The fixed points can be identified with ``memories'' stored in the weight matrices, and the entire neuron-astrocyte model can be regarded as an energy-based Dense Associative Memory \cite{krotov2016dense,krotov2023new}. Importantly, this framework allows us to show that the presence of a single astrocyte can provably boost the memory capacity per compute unit of a neural circuit by a factor of $N$.   

We will follow the general formulation of energy-based associative memories \cite{krotovLargeAssociativeMemory2021ICLR,krotovHierarchicalAssociativeMemory2021,hoover2022universal}, which starts with picking three Lagrangians, which define layers of our architecture (neurons, synapses, and astrocytic processes), and the corresponding activation functions. These Lagrangians are: a neural Lagrangian $\mathcal{L}^{[n]}$, a synaptic Lagrangian $\mathcal{L}^{[s]}$, and an astrocyte process Lagrangian $\mathcal{L}^{[p]}$. In general these scalar functions can be arbitrary (differentiable) functions of the corresponding dynamical variables. The details of our derivation can be found in Appendix \ref{section:lagrangian_definitions}.



From these Lagrangians, we can derive via a Legendre transformation three terms in the overall energy function of the neuron-astrocyte system: $E^{[n]}$, $E^{[s]}$, and $E^{[p]}$. The activation functions in our model are dictated by the Lagrangians--indeed, the $i^{th}$ activation is simply the partial derivative of the Lagrangian with respect to the $i^{th}$ dynamical variable (see Appendix \ref{section:lagrangian_definitions} and Figure \ref{figure:lagrangians}). The remaining contributions to the total energy of the system describe the interactions between neurons, synapses, and astrocytes. These contributions describe the synapse-mediated interactions between the neurons $E^{[ns]}$, the interactions between the processes and the synapses $E^{[ps]}$, and the interactions between the individual processes inside the astrocyte $E^{[pp]}$.

The overall energy function of the neuron-astrocyte model can now be written as the sum of these six terms
\begin{equation}
E = E^{[n]} + E^{[s]} + E^{[p]} + E^{[ns]} + E^{[ps]} + E^{[pp]} \label{eq:total energy}    
\end{equation}
From this Lagrangian formalism \cite{krotovLargeAssociativeMemory2021ICLR,krotovHierarchicalAssociativeMemory2021,hoover2022universal}, the dynamical equations for the associative neuron-astrocyte energy can be viewed as the negative gradient (with respect to the nonlinearities):
\begin{equation}
\begin{cases}
    \begin{split}
\tau_n \ \dot{x}_i &= -\frac{\partial E}{\partial \phi_i} = -\ \lambda \ x_i + \sum\limits^N_{j = 1} g_{ij} \phi_j\  \\
\tau_s \ \dot{s}_{ij} &= -2\frac{\partial E}{\partial g_{ij}} =-\ \alpha \ s_{ij} +  \phi_i\phi_j \ +\ \psi_{ij}\  \\
\tau_p \  \dot{p}_{ij} &= -2\frac{\partial E}{\partial \psi_{ij}} =  -\ \gamma \ p_{ij} + \sum\limits_{k,l = 1}^N  T_{ijkl} \ \psi_{kl} \ +\  g_{ij} \ 
    \end{split}
\end{cases}
\label{eq:neuron-astro-eqs}
\end{equation}
The energy-based equations have a large amount of symmetry--both in the parameters and the dynamical degrees of freedom, e.g., $T_{ijkl} = T_{klij}$,  see Appendix \ref{section:lagrangian_definitions}.  These symmetries, are needed for the existence of the global energy function for our neuron-astrocyte network, which leads to mathematical tractability. In real biology some (or all) of these symmetries might be broken, and the analytical tractability might be more difficult or even impossible. We use the energy-based model to establish theoretically the memory storage capabilities of our model. The non-symmetric model is studied numerically in section \ref{Section: Simulations} where we show that it possesses similar capabilities despite lacking the energy-based formulation. Note that unlike the symmetries in the original Hopfield networks \cite{hopfieldNeuralNetworksPhysical1982} (which have no known biological interpretation), the invariance of $\mathbf{T}$ with respect to swapping indices $ij$ and $kl$ can be viewed as a natural consequence of the underlying symmetry of calcium diffusion.

The first two equations in (\ref{eq:neuron-astro-eqs}) are reminiscent of the approach by Dong and Hopfield \cite{dong1992dynamic}, which describes both the neural dynamics and synaptic plasticity by a single energy function. The difference of our system compared to \cite{dong1992dynamic} is the existence of the network of astrocytic processes, which interact with each other and with the synapses.  Following the general Lagrangian formalism it can be shown (see appendix \ref{section:proof_of_energy}) that 
\begin{equation}\label{E_dot}
\begin{aligned}
    \frac{dE}{dt} \ = \ &- \bigg[ \tau_n \sum\limits_{i,j=1}^N \dot{x}_i \ \frac{\partial^2 \mathcal{L}^{[n]}}{\partial x_i \partial x_j} \ \dot{x}_j \ \\ 
    &\ + \ \ \frac{\tau_s}{2} \sum\limits_{i,j,k,l=1}^N \dot{s}_{ij} \ \frac{\partial^2 \mathcal{L}^{[s]}}{\partial s_{ij} \partial s_{kl}} \ \dot{s}_{kl}\ \\
    & \ + \ \frac{\tau_p}{2} \sum\limits_{i,j,k,l=1}^N \dot{p}_{ij} \ \frac{\partial^2 \mathcal{L}^{[p]}}{\partial p_{ij} \partial p_{kl}} \ \dot{p}_{kl} \ \bigg] \\
    &\leq \  0 
\end{aligned}
\end{equation}
The last equality sign holds if each Lagrangian has a positive semidefinite Hessian matrix. When the Hessian matrices are strictly positive definite, the dynamical equations (\ref{eq:neuron-astro-eqs}) are guaranteed to arrive at a fixed point, because the energy is bounded from below (through the invariant set theorem~\cite{slotine1991applied}). Thus, starting from an initial state the network dynamics flows towards one of the fixed points and for this reason describes the operation of an associative memory.

\begin{figure}[ht!]
  \centering
  \includegraphics[width=\textwidth]{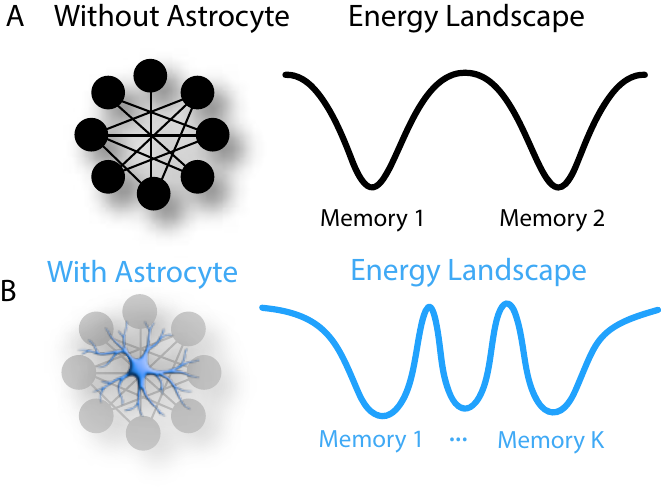}
  \caption{For a fixed number of neurons, the neuron-astrocyte network is capable of storing many more memories than the neuron-only network. A) The energy landscape of a neuron-only associative network. B) The energy landscape of a neuron-astrocyte associative network. The memories are more densely packed into the state-space, thereby enabling superior memory storage and retrieval capabilities.}
  \label{fig:energy_landscape}
\end{figure}
\vspace{-0.3cm}
\subsection{Connection to Dense Associative Memory}
Energy-based neuron-astrocyte networks are described by a sophisticated system of nonlinear differential equations (\ref{eq:neuron-astro-eqs}) that are guaranteed to represent dynamical trajectories converging to fixed points attractors, assuming certain conditions on the Lagrangians are met. The locations of those fixed points $x_i^*$, $s_{ij}^*$, and $p_{ij}^*$ coincide with the local minima of the energy function \eqref{eq:total energy}, and are independent of the time scales $\tau_n$, $\tau_s$, and $\tau_p$. The kinetics of the model (i.e., the shape of dynamical trajectories), however,  heavily depends on these time scales. Although the characteristic time scales of synaptic plasticity and dynamics of processes (as well as the dynamics of the entire astrocyte) are subjects of active debates in the community, it is generally believed that neurons operate on faster time scales than synaptic plasticity or the processes, $\tau_n\ll\tau_s, \tau_p$. Since the goal of this section is to analyse the fixed points of this network, which are independent of these time scales, we have a freedom to choose the kinetic time scales in a way dictated by mathematical convenience, rather than biological reality. Specifically, we will derive an effective dynamics on neurons that arises after the synapses and astrocytes are integrated out from the dynamical equations, which intuitively\footnote{Strictly speaking, specific conditions must be satisfied in order for this operation to be well-defined and to avoid pathologies such as peaking phenomena. In our system this can easily be done by restricting the eigenvalues of $\mathbf{T}$ to be less than $\gamma$, which ensures that the fast dynamics are contracting ~\cite{delvecchioContractionTheoryApproach2013}.} is possible if $\tau_s, \tau_p\ll\tau_n$. 
Despite this ``unbiological'' choice for the intermediate steps, the final answer represents accurate locations of fixed points for the network that operates in the ``biological'' regime $\tau_n\ll\tau_s, \tau_p$, or with any other choice of time scales.   

The fixed points of the synaptic and processes' dynamics in \eqref{eq:neuron-astro-eqs} are defined by (assuming for simplicity that $\alpha = \gamma = 0$)
\begin{equation*}
    \begin{cases}
        \psi_{ij} = -\phi_i\phi_j\\
        g_{ij} = \sum\limits_{k,l=1}^N T_{ijkl} \phi_k \phi_l
    \end{cases}
\end{equation*}
Note that for a fixed value of $\phi_i$, these equations uniquely determine the values of $s_{ij}$ and $p_{ij}$ when $g$ and $\psi$ are strictly monotonic. Substituting this solution into the first equation (\ref{eq:neuron-astro-eqs}), gives the effective dynamics for the neural dynamics (assuming for simplicity that $\lambda=1$)
\begin{equation}
    \tau_n \dot{x}_i =   - x_i + \sum\limits_{j,k,l=1}^N T_{ijkl}\ \phi_j\phi_k\phi_l \label{effective eq}
\end{equation}
and effective energy 
\begin{equation}
    E^\text{eff} = \Big[\sum_{i = 1}^N x_i\phi_i - \ \mathcal{L}^{[n]} \Big] - \frac{1}{4}\sum\limits_{i,j,k,l=1}^N T_{ijkl}\  \phi_i\phi_j\phi_k\phi_l \label{effective energy}
\end{equation}
Equations (\ref{effective eq},\ref{effective energy}) contain the essence of our theoretical argument. The fixed points of this effective dynamical system exactly coincide with the fixed points of the original complete energy-based neuron-astrocyte network (\ref{eq:neuron-astro-eqs}), projected on the neuron-only subspace. The hallmark of this effective theory is the existence of the four-body neuron-to-neuron interactions, represented by the product of four firing rate functions $\phi_i$ in the effective Lyapunov function and the product of three firing rate functions in the effective equations. In conventional firing rate models there is only one firing rate function in the right hand side of the dynamical equations and two firing rate functions in the corresponding energy function, see for example \cite{hopfield1984neurons}. This is a mathematical reflection of the biological fact that each synapse connects two neurons (pre- and post-synaptic cells). In our model we have demonstrated that the contribution of the astrocyte is to effectively create a computational many-neuron synapse (mediated by the network of astrocytic processes). In other words, the computational function of the astrocyte is to bring the information about the states of distant synapses (and neurons) to each tripartite synapse resulting in the ``effective'' four-neuron synapse that connects neurons that are potentially very far away from each other. In what follows we will explain that this computational property has important implications for storing memories. 

\begin{figure*}[ht!]
  \centering
  \includegraphics[width=\textwidth]{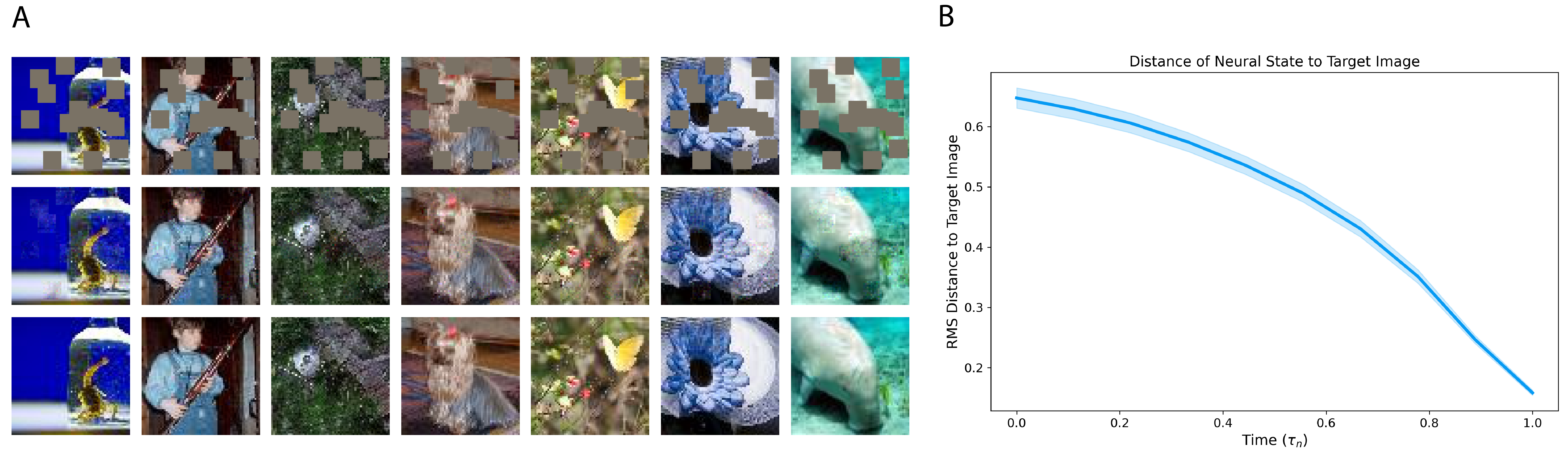}
  \caption{A) Error-correcting capabilities of the neuron-astrocyte network, trained with backpropagation, demonstrated with images from the Tiny ImageNet dataset \cite{wu2017tiny}. Top row is the masked input to the network, middle row is the final state of our network, bottom row is the ground-truth, unmasked image. B) Root-mean-squared distance of the state of our network to the ground-truth image, as function of time. Standard error was calculated across a batch of 64 images.}
  \label{fig:tiny_imagenet}
\end{figure*}

\subsection{Storing Memories in the \\ Network of Astrocyte Processes}
Imagine that we are given $K$ memory patterns $\mathbf{\xi}^\mu$ (index $\mu=1...K$), and each pattern is an $N$-dimensional vector. The task of associative memory is to store these patterns in the weights of the neural-astrocyte network, so that temporal dynamics can asymptotically flow to these patterns. Choose tensor $\mathbf{T}$ such that it satisfies the following relationship
\begin{equation}
    T_{ijkl} \equiv \sum_{\mu=1}^K \xi^{\mu}_i \ \xi^{\mu}_j \ \xi^{\mu}_k \ \xi^{\mu}_l \label{eq: definition of T}
\end{equation}
With these notations the effective neuron-only theory is equivalent to a model with quartic interaction from the Dense Associative Memory family \cite{krotov2016dense, krotov2023new}. Specifically, the effective energy can be written as
\begin{align*}
    E^\text{eff} = \Big[\sum_{i = 1}^N x_i\phi_i - \ \mathcal{L}^{[n]} \Big] - \sum_{\mu=1}^K F\Big(\sum_{i=1}^N \xi^\mu_i \phi_i \Big), \hspace{0.5cm} \\ \text{where} \hspace{0.5cm} F(z) = \frac{1}{4}z^4 \label{effective energy with mem}
\end{align*}
and effective equations as 
\begin{equation}
    \tau_n \dot{x}_i = - x_i + \sum_{\mu=1}^K \xi^\mu_i F^{\prime}\Big(\sum_{j=1}^N \xi^\mu_j \phi_j \Big)  \label{effective eq with mem}
\end{equation}
Dense Associative Memories is a new class of models that extend traditional Hopfield Networks \cite{hopfieldNeuralNetworksPhysical1982,hopfield1984neurons} by introducing higher than quadratic terms in their energy function. It has been shown that this extension leads to superior information storage capacity and representational power of these models compared to the traditional Hopfield Networks \cite{krotov2016dense}. They are also related to the attention mechanism in Transformers \cite{ramsauer2020hopfield,krotovLargeAssociativeMemory2021ICLR} and are used in state-of-the-art, energy-based neural networks \cite{hoover2024energy, hoover2023memory}. 

\vspace{-0.2cm}
\paragraph{Memory Capacity of a Neuron-Astrocyte Network} It is insightful to ask the question: how many memories can the model store \textit{per the number of compute units}? Assuming a conservative definition\footnote{A less conservative definition would consider the entire astrocyte as a single compute unit.} of the ``compute unit'', our energy-based neuron-astrocyte model (\ref{eq:neuron-astro-eqs}) has approximately $N^2$ compute units in the limit of large $N$, i.e. 
\begin{equation*}
     N\ \text{neurons} + N^2\ \text{synapses} + N^2\ \text{processes} \sim N^2\ \text{compute units}
\end{equation*}
The storage capacity $K^\text{max}$ of the Dense Associative Memory model with quartic energy is known to be \cite{krotov2016dense}
\begin{equation*}
\hspace{1.5cm} K^\text{max} \ \sim \ N^3
\end{equation*}
Thus, for our model the number of memories per compute unit grows linearly as the size of the network is increased
\begin{equation*}
    \frac{K^\text{max}}{\text{Number of compute units}} \ \sim \ N
\end{equation*}
This metric can be compared with other biologically-plausible implementations of Dense Associative Memory. For instance Krotov and Hopfield \cite{krotovLargeAssociativeMemory2021ICLR}  have proposed to augment the network of feature neurons with a set of auxiliary hidden neurons. In their model, the compute units consist of both feature and hidden neurons, and importantly the number of memories per compute unit is a constant independent of $N$,
\begin{equation*}
    \frac{K^\text{max}}{\text{Number of compute units}} \ \sim \ constant
\end{equation*}
Thus, neuron-astrocyte networks significantly outperform the implementation \cite{krotovLargeAssociativeMemory2021ICLR} according to this metric, see Figure \ref{fig:energy_landscape}. This makes neuron-astrocyte networks an exciting candidate for biological ``hardware'' implementing Dense Associative Memory.

Even more insightful is to trace down the origin of this memory storage back to the biophysical implementation of the neuron-astrocyte networks \eqref{eq:neuron-astro-eqs}. The memories in our model are stored in tensor $\mathbf{T}$, which describes the network of astrocyte's processes and the transport of \ca or other potential molecules, e.g. protein kinase A, between those processes. Our theory suggests that memories can be stored in the biological machinery (inside a single astrocyte) of transport of an appropriate signalling molecule between the astrocyte's processes. We have demonstrated a conceptual theoretical possibility of such a storage through a Hebbian-like plasticity rule (\ref{eq: definition of T}). This was done for the sake of transparency of the theoretical argument. In principle, more sophisticated storage rules are possible too. We hope that  future experiments might be able to test this exciting hypothesis. 
\vspace{-0.2cm}
\paragraph{How Many Astrocyte Parameters are Needed?} 
The Hebbian-like storage scheme \eqref{eq: definition of T} requires that every process within the astrocyte be directly connected to every other process. Future experiments will hopefully determine if such detailed communication within a single astrocyte is possible. In the meantime, it is instructive to ask how the memory capacity changes as a function of the degree of process-to-process connectivity. 

Heuristically, if we wish to store $K$ memories, each containing $N$ independent bits, we need on the order of $KN$ parameters in our model--one parameter for each bit of information stored. For example, if we wish to store $K = N$ memories, we need on the order of $N^2$ parameters--these parameters can be stored inside the $N \times N$ weight matrix of a traditional Hopfield network. For our neuron-astrocyte network, the number of parameters can be written as $rN^2$, where $r$ is the number of connections per astrocyte process. For an all-to-all network as we have described above, $r = N^2$. Thus, the number of connections per process needed to store $K$ memories is given by
\[KN = rN^2 \quad \implies \quad r = \frac{K}{N}\]
This equation tells us that if we wish to achieve linear storage capacity (i.e., K = N), then we can ignore process-to-process connectivity, since $r = const$ means that each astrocyte process is an isolated dynamical variable. If we wish to achieve supralinear memory storage, one way to do so is by connecting the astrocyte processes to one another. Biologically, parameter $r$ can be larger than constant, but smaller than $N^2$. Thus, this equation provides the number of memories that can be stored \textit{given} a particular value of $r$ -- a quantity which can in principle be determined experimentally. 
\vspace{-0.2cm}
\paragraph{Connection to Transformers Models}
It is worth noting that having detailed entries of the tensor $T_{ijkl}$ is not necessary for our neuron-astrocyte model to perform interesting computations. Indeed, one can demonstrate (see Appendix \ref{section:proof_of_transformer}) that setting $T_{ijkl} = 1$ in equation \eqref{eq:astro_eqs} results in a stable dynamical model whose equilibrium states approximate the output of a transformer's self-attention mechanism \cite{kozachkov2023building}.

\vspace{-0.3cm}
\section{Simulations}\label{Section: Simulations}
In this section we conduct two computational experiments. The first uses the energy-based equations \eqref{eq:neuron-astro-eqs} with the Hebbian-like learning rule \eqref{eq: definition of T}. The second employs backpropagation-through-time, foregoing symmetry requirements. The aim of the first experiment is to validate our theoretical claims. The second experiment aims to demonstrate that strong symmetry is not necessary, but rather sufficient, for the system to exhibit associative memory function. This point is crucial in the context of biology, considering the difficulty of achieving `pure' symmetry on noisy biological hardware.
\vspace{-0.3cm}
\paragraph{Energy-Based Experiments}
To demonstrate that the memory storage scheme \eqref{eq: definition of T} above works in practice, we performed numerical experiments using the CIFAR10 dataset. Figure \ref{fig:cifar} shows the result of retrieving four different memories after encoding $K = 25$ memories in the network. As predicted by theory, in each case the neuron-astrocyte network converges to fixed points with the correct neural attractor (i.e., the attractor corresponding to the stored memory). Along trajectories of the neuron-astrocyte network, the energy function is monotonically decreasing.  The details of the training are given in Appendix \ref{section:astro_energy_based}.
\vspace{-0.2cm}
\paragraph{Backpropagation-Based Experiments}
In order to show that our network does not require large amounts of symmetry to perform associative memory functions, we trained it on a self-supervised learning task using the Tiny ImageNet dataset \cite{wu2017tiny}. Specifically, given a batch of ImageNet images, downsampled to $64 \times 64$, we randomly masked fifteen square patches, each ten pixels across (roughly 40\% of all pixels). The exact positions of the mask patches was varied across batches. The masked images were provided to the network as an initial state, and then the state of the network was dynamically evolved for prespecified number of steps. The goal of training was to minimize the difference (as measured by least squares) between the network output and the unmasked images. The parameters of the network were initialized randomly and optimized using Backpropagation-Through-Time (BPTT). The results are shown in Figure \ref{fig:tiny_imagenet}. More details on the training process can be found in Appendix \ref{section:astro_backprop}. 

Finally, we note that the energy-based model can also be trained using BPTT. Indeed, due to fixed-point convergence properties, it can be trained using implicit techniques such such as recurrent backpropagation \cite{pineda1987generalization,bai2019deep}.
\vspace{-0.3cm}
\section{Discussion}
We have introduced a biologically-inspired model that describes the interactions between neurons, synapses, and astrocytes. In our model, astrocytes are able to adaptively control synaptic weights in an online fashion. Theoretical analysis has demonstrated that this model can exhibit associative memory and is closely related to the Dense Associative Memory family of models with supralinear memory capacity. Furthermore, we have presented a simple algorithm for memory storage and have provided numerical evidence of its effectiveness, such as successfully storing and retrieving CIFAR10 and ImageNet images. 

In broader terms, this work proposes that memories can, at least in part, be stored within the molecular machinery of astrocytes. This contrasts with the prevailing neuroscience viewpoint that memories are stored in the synaptic weights between neurons. To experimentally validate this claim, one would need to selectively interfere with the ability of \ca to diffuse intracellularly through astrocytes. Our model predicts that hindering this diffusion would significantly impair memory recall. While our focus has been on a mini-circuit consisting of a single astrocyte interacting with multiple nearby synapses, astrocytes also extensively communicate with each other through chemical gap junctions. Exploring the implications of this intercellular coupling will be the subject of future research. 

Key ideas in machine learning and AI drew initial inspiration from neuroscience, including neural networks, convolutional nets, threshold linear (ReLu) units, and dropout. Yet it is debatable whether neuroscience research from the last fifty years has significantly influenced or informed machine learning. Astrocytes, along with other biological structures such as dendrites \cite{boahen2022dendrocentric}, may offer a fresh source of inspiration for building state-of-the-art AI systems.

\paragraph{Author contributions}
L.K \& D.K conceptualized the project, L.K, J.J.S \& D.K derived the results and wrote the paper. 
\paragraph{Competing interests}
The authors declare no competing interests.
\paragraph{Computer Code}
The code used in this paper is available at \texttt{https://github.com/kozleo/naam}.  

\bibliography{biblio1,biblio2}

\newpage
\appendix

\section{Definitions of Lagrangians and Energy}\label{section:lagrangian_definitions}

As described in the main text, the Lagrangians are: a neural Lagrangian $\mathcal{L}^{[n]}$, a synaptic Lagrangian $\mathcal{L}^{[s]}$, and an astrocyte process Lagrangian $\mathcal{L}^{[p]}$. In general these scalar functions can be arbitrary (differentiable) functions of the corresponding dynamical variables. The activation functions are defined as partial derivatives of the Lagrangians 
\begin{equation}\label{eq:lagrangians}
\underbrace{\mathcal{L}^{[n]}(\mathbf{x}) \ \ \rightarrow \ \ \phi_i \equiv \pp{\mathcal{L}^{[n]}}{x_i}}_{\text{Neural Lagrangian}}, \hspace{1cm} \underbrace{\mathcal{L}^{[s]}(\mathbf{s}) \ \ \rightarrow \ \ g_{ij} \equiv \pp{\mathcal{L}^{[s]}}{s_{ij}}}_{\text{Synaptic Lagrangian}}, \hspace{1cm} \underbrace{\mathcal{L}^{[p]}(\mathbf{p}) \ \ \rightarrow \ \ \psi_{ij} \equiv \pp{\mathcal{L}^{[p]}}{p_{ij}}}_{\text{Astrocyte Process Lagrangian}} 
\end{equation}
One possible choice of these functions is additive: summing each contribution from all the individual computational elements (e.g., individual neurons), which results in activation functions that depend only on individual computational elements -- for instance, $\phi(x_i) = \text{tanh}(x_i)$. More general choices of the Lagrangians allow for ``collective" activation functions, which depend on the dynamical degrees of freedom of several or all the computational elements in a given layer, for example a softmax.

From the Lagrangians \eqref{eq:lagrangians}, we may derive via a Legendre transform three terms in the overall energy function of the neuron-astrocyte system,  corresponding to three layer energies,
\begin{equation}\label{eq:neuron_process_energy}
E^{[n]}+E^{[s]}+E^{[p]} = \underbrace{\lambda \bigg[\sum_{i = 1}^N \ x_i\phi_i - \mathcal{L}^{[n]} \bigg]}_{\text{Neural Energy}} \ +\ \underbrace{ \frac{\alpha}{2}\bigg[\sum_{i,j = 1}^N \ s_{ij}g_{ij} - \mathcal{L}^{[s]} \bigg]}_{\text{Synaptic Energy}} \ +\  \underbrace{\frac{\gamma}{2} \bigg[\sum_{i,j = 1}^N \ p_{ij}\psi_{ij} - \mathcal{L}^{[p]} \bigg]}_{\text{Astrocyte Process Energy}}    
\end{equation}
where for simplicity of the presentation we dropped the input signals, $b_i = c_{ij} = d_{ij} = 0$. The remaining contributions to the total energy of the system describe the interactions between these three layers. These contributions describe the synapse-mediated interactions between the neurons $E^{[ns]}$, the interactions between the processes and the synapses $E^{[ps]}$, and the interactions between the individual processes inside the astrocyte $E^{[pp]}$, 
\begin{equation}\label{eq:interaction_energy}
E^{[ns]} + E^{[ps]} + E^{[pp]} = - \bigg[ \frac{1}{2}\sum_{i,j = 1}^N g_{ij}(\mathbf{s})\phi_i(\mathbf{x}) \phi_j(\mathbf{x})\ +\ \frac{1}{2}\sum_{i,j = 1}^N \psi_{ij}(\mathbf{p}) g_{ij}(\mathbf{s})\ +\ \frac{1}{4}\sum_{i,j,k,l = 1}^N T_{ijkl} \ \psi_{ij}(\mathbf{p}) \psi_{kl}(\mathbf{p}) \bigg]  
\end{equation}
The overall energy function of the neuron-synapse-astrocyte model can now be written as the sum of these six terms
\begin{equation}
E = E^{[n]} + E^{[s]} + E^{[p]} + E^{[ns]} + E^{[ps]} + E^{[pp]} \label{total energy}    
\end{equation}
As mentioned previously, the energy-based equations have a large amount of symmetry--both in the parameters and the dynamical degrees of freedom. Specifically, $ \ s_{ij}=s_{ji}, \ g_{ij}=g_{ji}, \ p_{ij}=p_{ji}, \ \psi_{ij}=\psi_{ji}$, and $T_{ijkl} = T_{klij},\ T_{ijkl}=T_{jikl},\ T_{ijkl} = T_{ijlk}$. These symmetries, are needed for the existence of the global energy function for our neuron-astrocyte network, which leads to mathematical tractability. In real biology some (or all) of these symmetries might be broken, and the analytical tractability might be more difficult or even impossible. We use the energy-based model to establish theoretically the memory storage capabilities of our model. The non-symmetric model is studied numerically in section \ref{Section: Simulations}.

\section{Proof of Decreasing Energy Function}\label{section:proof_of_energy}
The overall time derivative of the energy function may be written as
\[\frac{dE}{dt} = \sum_{i = 1}^N \ \pp{E}{x_i} \ \dot{x}_i + \sum_{i,j = 1}^N \ \pp{E}{s_{ij}} \ \dot{s}_{ij} + \sum_{i,j = 1}^N \ \pp{E}{p_{ij}} \ \ \dot{p}_{ij} \]
which may be expressed using the chain rule as
\begin{equation}
\begin{aligned}
\frac{dE}{dt} &= \sum_{i,j = 1}^N \ \pp{E}{\phi_i}\ \pp{\phi_i}{x_j} \ \dot{x}_j + \sum_{i,j,k,l = 1}^N \ \pp{E}{g_{ij}}\pp{g_{ij}}{s_{kl}} \ \dot{s}_{kl} + \sum_{i,j,k,l = 1}^N \ \pp{E}{\psi_{ij}}\pp{\psi_{ij}}{p_{kl}} \ \dot{p}_{kl}  \\ 
&= \sum_{i,j = 1}^N \ \pp{E}{\phi_i}\ \frac{\partial^2 \mathcal{L}^{[n]}}{\partial x_i \partial x_j} \ \dot{x}_j + \sum_{i,j,k,l = 1}^N \ \pp{E}{g_{ij}}\frac{\partial^2 \mathcal{L}^{[s]}}{\partial s_{ij} \partial s_{kl}} \ \dot{s}_{kl} + \sum_{i,j,k,l = 1}^N \ \pp{E}{\psi_{ij}}\frac{\partial^2 \mathcal{L}^{[p]}}{\partial p_{ij} \partial p_{kl}} \ \dot{p}_{kl} 
\end{aligned}
\end{equation}
The second line follows from the definition of the Lagrangians \eqref{eq:lagrangians}. Plugging the dynamics defined in equations \eqref{eq:neuron-astro-eqs} into this last expression, we get the desired result, provided that the Lagrangians are all convex (i.e., have positive semi-definite Hessians)
\begin{equation}
 \frac{dE}{dt} = - \bigg[ \tau_n \sum_{i,j = 1}^N \ \dot{x}_i \ \frac{\partial^2 \mathcal{L}^{[n]}}{\partial x_i \partial x_j} \ \dot{x}_j \ + \ \frac{\tau_s}{2} \sum_{i,j,k,l = 1}^N \ \dot{s}_{ij} \ \frac{\partial^2 \mathcal{L}^{[s]}}{\partial s_{ij} \ \partial s_{kl}} \ \dot{s}_{kl} \ + \ \frac{\tau_p}{2} \sum_{i,j,k,l = 1}^N \ \dot{p}_{ij} \ \frac{\partial^2 \mathcal{L}^{[p]}}{\partial p_{ij} \partial p_{kl}} \ \dot{p}_{kl} \bigg] \leq 0     
\end{equation}
\section{Proof of Neuron-Astrocyte Equilibration to Transformer Output}\label{section:proof_of_transformer}
\begin{figure*}[ht!]
  \centering
  \includegraphics[width=0.5\textwidth]{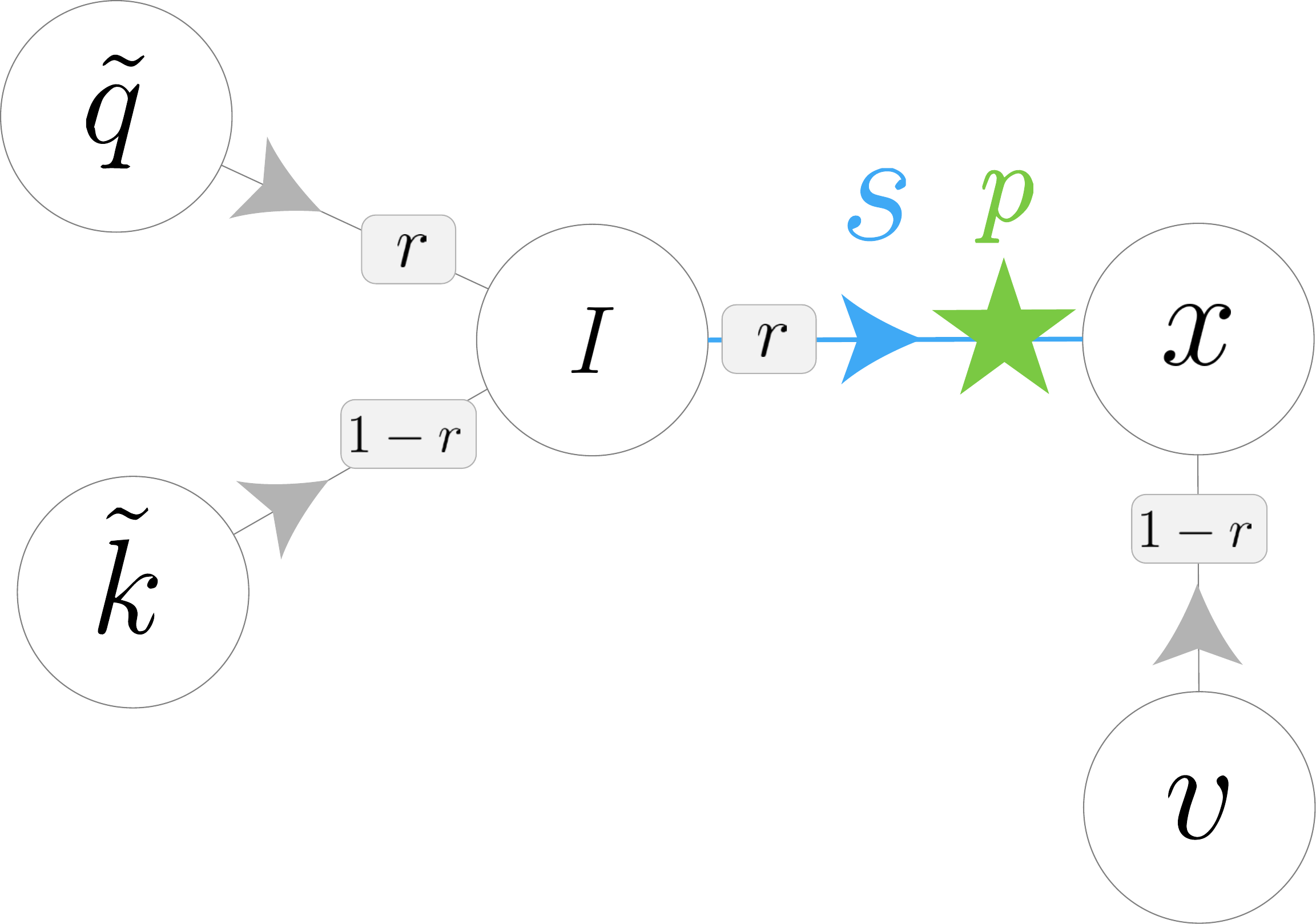}
  \caption{Dynamic, stable neuron-astrocyte architecture which implements the self-attention operation in transformers.}
  \label{fig:transformer_nature}
\end{figure*}

\paragraph{Neuron-Astrocyte Transformer Architecture}
The aim of this section is to demonstrate that a simple selection of the astrocyte process-to-process weights \(T_{ijkl} = 1\) is sufficient, along with a specialized architecture (Figure \ref{fig:transformer_nature}), to produce interesting computations in the general neuron-astrocyte network equations \eqref{eq:neuron_eqs}, \eqref{eq:synaptic_dynamics}, \eqref{eq:astro_eqs}. We consider a single group of \(N\) neurons, where the state of the \(i\)-th neuron in this group is denoted by \(x_i\). These neurons receive inputs from another group of \(M\) neurons, where the state of the \(j\)-th neuron in this group is denoted \(I_j\). The synaptic connection between neuron \(I_j\) and neuron \(x_i\) is represented by \(s_{ij}\). The $x_i$ neurons also receive input from another group of $N$ neurons, whose state we denote by $v_i$, for reasons that will become clear later on. The dynamical equations for the \(x_i\) layer are given by
\begin{equation}\label{eq:neuron_trans}
\tau_n \dot{x}_i = -x_i + r\sum_{j = 1}^M s_{ij} I_j + (1-r)v_i
\end{equation}
where $r = \{0,1\}$ stands for "read", and is a global parameter controlling whether the network is in "read" or "write" mode. Biologically, global coordination of this kind may be achieved by neuromodulators (e.g., acetylcholine) \cite{tyulmankovBiologicalLearningKeyvalue2021}. We additionally assume that the $I_j$ neurons receive strong input from two $M$-dimensional neural populations which we denote as $\tilde{q}_j$ and $\tilde{k}_j$ (again for reasons that will become clear shortly), so that the state of neuron $I_j$ is given by
\begin{equation}\label{eq:I_eqs}
I_j = r \ \tilde{q}_j + (1-r) \ \tilde{k}_j     
\end{equation}
The synaptic weights $s_{ij}$ are modulated by an astrocyte and evolve according to the following dynamical equations:
\begin{align}\label{eq:synapse_trans}
\tau_s \dot{s}_{ij} = -p_{ij} s_{ij} + c_{ij}
\end{align}
where \(p_{ij}\) represents the state of the astrocyte process \(ij\), and \(c_{ij}\) is a fixed bias term. This set of synaptic equations can be associated with equations \eqref{eq:synaptic_dynamics} by setting
\[ \alpha = 0, \quad \text{and} \quad f(s_{ij}, x_i, x_j, p_{ij}) = -p_{ij} s_{ij} \]
The astrocyte dynamics are described by simple diffusive equations:
\begin{align}\label{eq:astro_trans}
\tau_p \dot{p}_{ij} = \sum_{k = 1}^N \sum_{l = 1}^M \left[p_{kl} - p_{ij}\right] \quad \text{with} \quad \sum_{i = 1}^N \sum_{j = 1}^M p_{ij}(0) > 0 
\end{align}
The inequality is to ensure that the total amount of \ca initially in the astrocyte is positive. Biologically, even \ca concentrations inside individual processes are positive $p_{ij}(0)\geq0$, but, mathematically, we will only use the positivity of the total amount of calcium inside the astrocyte. Similar to the synaptic variables, this set of astrocyte equations can be associated with the astrocyte equations \eqref{eq:astro_eqs} by setting
\[\psi(p_{ij}) = p_{ij}, \quad \gamma = NM, \quad T_{ijkl} = 1, \quad \kappa(s_{ij}) = 0 \quad \text{and} \quad d_{ij} = 0 \]
Before establishing a connection with transformer networks, we will describe the dynamical properties of Equations \eqref{eq:neuron_trans}, \eqref{eq:synapse_trans}, and \eqref{eq:astro_trans}. Specifically, we will demonstrate that, during the reading phase, the neurons $x_i$ converge to an equilibrium point determined solely by the input neurons $I_j$, the initial \ca concentration in the astrocyte, and the synaptic bias terms $c_{ij}$. Following this, we will illustrate how a judicious and biologically plausible selection of input neuron states, initial \ca levels, and synaptic biases enables the neurons $x_i$ to mimic the output of the self-attention mechanism in transformers.

\paragraph{Convergence \& Synchronization of Astrocyte Processes}
To begin, note that the astrocyte equations \eqref{eq:astro_trans} are autonomous with respect to the neural and synaptic variables. Therefore, we can analyze their convergence properties independently from these variables.
In particular, we can show that the astrocyte equations synchronize to the average of their initial conditions. To see this, first note that the total amount of \ca in the astrocyte, which we denoted $z$, is conserved throughout the diffusion process
\[z \equiv \sum_{i = 1}^N \sum_{j = 1}^M  p_{ij} \hspace{0.5cm} \implies \hspace{0.5cm} \dot{z} = \sum_{i = 1}^N \sum_{j = 1}^M  \dot{p}_{ij} = 0 \]
Second, note that this property implies that if the astrocyte processes \textit{synchronize}, i.e., $p_{ij} = p_{kl} = p^*$, then the state of each astrocyte process must converge to the average of the astrocyte initial conditions, because
\begin{equation}\label{eq:p_trans_fp}
z(t)  \ = \ NM p^* \ = \ z(0)  \ = \  \sum_{i = 1}^N \sum_{j = 1}^M  p_{ij}(0) \hspace{0.5cm} \implies  \hspace{0.5cm} p^* = \frac{1}{NM}\sum_{i = 1}^N \sum_{j = 1}^M  p_{ij}(0) \ > \ 0 
\end{equation}
The inequality follows from the assumption in \eqref{eq:astro_trans}, that the total initial amount of \ca in the astrocyte is positive.  To prove that the astrocyte processes in fact synchronize, one can use a virtual system, as in  \cite{wang2005partial} or a Lyapunov-like function
\[L = \frac{1}{2} \ (p_{ij} - p_{kl})^2 \geq 0 \]
for arbitrary indices $ij$ and $kl$. Taking the time derivative of this function, one sees that 
\begin{align*}
\dot{L} = (p_{ij} - p_{kl})(\dot{p}_{ij} - \dot{p}_{kl}) = -\frac{NM}{\tau_p}(p_{ij} - p_{kl})^2 = -\frac{2NM}{\tau_p} L \quad \implies \quad L(t) = L(0) e^{-\frac{2NMt}{\tau_p}}
\end{align*}
which shows that the astrocyte processes do in fact synchronize (i.e., $|p_{ij} - p_{kl}| \rightarrow 0$) exponentially with rate $\frac{NM}{\tau_p}$.
\paragraph{Convergence of Synapses}
Moving on to the synaptic equations \eqref{eq:synapse_trans}, we will assume that the astrocyte processes have converged to $p^* > 0$. This assumption is justified because, as the preceding paragraph shows, the converge of the astrocyte process to $p^*$ is \textit{exponential}, meaning that $p_{ij}$ can be brought arbitrarily close to $p^*$ after finite time. Because $c_{ij}$ is a constant, and because $p^*$ is strictly positive, this implies that the synapses simply converge exponentially quickly to the value
\begin{equation}\label{eq:s_trans_fp}
s^*_{ij} = \frac{c_{ij}}{p^*}    
\end{equation}
\paragraph{Convergence of Neurons}
Following a similar logic, the neural equations \eqref{eq:neuron_trans} converge exponentially. When the network is in its writing phase (i.e., $r = 0$), the neurons converge to the equilibrium point
\begin{equation}
x^*_i =  v_i   
\end{equation}
otherwise, when the network is in the reading phase (i.e., $r = 1$), the network converges exponentially to the equilibrium point
\begin{equation}\label{eq:x_trans_fp}
x^*_i =  \sum_{j = 1}^M s^*_{ij} I_j = \frac{1}{p^*} \sum_{j = 1}^M c_{ij}I_j = \frac{ NM \sum\limits_{j = 1}^Mc_{ij}I_j}{\sum\limits_{i = 1}^N \sum\limits_{j = 1}^M  p_{ij}(0)}   
\end{equation}
The first equality was obtained by substituting in $s^*_{ij}$ from \eqref{eq:s_trans_fp}, while the second equality was obtained by subsituting in the value of $p^*$ from \eqref{eq:p_trans_fp}. 

\paragraph{Transformer Self-Attention}
We are now in a position to relate the neural fixed point \eqref{eq:x_trans_fp} to the output of the self-attention mechanism in transformers. To establish this connection, we define several important terms. Consider a set of $K_{\text{tok}}$ \textit{tokens}, which are vectors in $\mathbb{R}^D$. As is standard in transformer architectures, these tokens are transformed via three linear mappings into three new sets of vectors known as keys, queries, and values. By collecting these transformed vectors into matrices, we denote
\[K, \ Q \in \ \mathbb{R}^{K_{\text{tok}} \times D} \qquad \text{and} \qquad  V \ \in \ \mathbb{R}^{K_{\text{tok}} \times N}.\]
The \textit{self-attention} matrix $A$ associated with these matrices is given by
\[A_{\mu i} = \sum\limits_{\beta=1}^{K_\text{tok}} \frac{\exp\left(\sum\limits_{s=1}^D Q_{\mu s} K_{\beta s}\right) V_{\beta i}}{\sum\limits_{\sigma=1}^{K_\text{tok}} \exp\left(\sum\limits_{s=1}^D Q_{\mu s} K_{\sigma s}\right)}\]
An important characteristic of the above self-attention matrix is that it may be approximated via feature maps \cite{rahimi2007random} with the following property
\[\phi(\mathbf{x})^T\phi(\mathbf{y}) \approx \text{exp}(\mathbf{x}^T\mathbf{y}) \]
where $\mathbf{x}$ and $\mathbf{y}$ are two vectors. In general, the output dimension of $\phi$, which we denote $M$ (the same $M$ as above) is much larger than the input dimension $D$. To keep notations clean, we define the output of these feature maps (applied column-wise to the matrices K and Q) as
\[  \tilde{K},\tilde{Q} \ \equiv \ \phi(K), \phi(Q) \ \ \in \ \ \fR^{K_{tok} \times M}\]
With this notation, we have that 
\[A_{\mu i} \approx \sum\limits_{\beta=1}^{K_\text{tok}} \frac{\sum\limits_{j=1}^M \tilde{Q}_{\mu j} \tilde{K}_{\beta j} V_{\beta i}}{\sum\limits_{\sigma=1}^{K_\text{tok}} \sum\limits_{j=1}^M \tilde{Q}_{\mu j} \tilde{K}_{\sigma j}}\]
\paragraph{Neuron-Astrocyte Self-Attention}
To make a connection to the fixed point equation \eqref{eq:x_trans_fp}, we first rearrange the above terms as follows
\begin{equation}\label{eq:approx_self_attn}
A_{\mu i} \approx \frac{\sum\limits_{j=1}^M \left(\sum\limits_{\beta=1}^{K_\text{tok}} V_{\beta i} \tilde{K}_{\beta j}\right) \tilde{Q}_{\mu j}}{\sum\limits_{j=1}^M \tilde{Q}_{\mu j} \sum\limits_{\sigma=1}^{K_\text{tok}} \tilde{K}_{\sigma j}}
\end{equation}
We then set the bias terms \(c_{ij}\) in the synaptic equations as follows:
\begin{equation}\label{eq:neuron_astro_trans_C}
c_{ij} = \frac{1}{M} \sum_{\beta = 1}^{K_{\text{tok}}} V_{\beta i} \tilde{K}_{\beta j}
\end{equation}
Biologically, this corresponds to a simple form of Hebbian learning between two groups of neurons. Within the framework of \eqref{eq:x_trans_fp}, this can be achieved during the writing phase (i.e., \(r = 0\)), such that \(x_i = v_i = V_{\beta i}\) and \(I_j = \tilde{k}_j \equiv \tilde{K}_{\mu j}\) (from \eqref{eq:I_eqs}). Then, updating \(c_{ij}\) by adding the product of these two terms for each \(\beta\) represents a simple form of associative Hebbian learning, and yields \eqref{eq:neuron_astro_trans_C}. Assuming \(c_{ij}\) is initially zero, we see that
\[
\Delta c_{ij} = \frac{1}{M} x_i I_j = \frac{1}{M} V_{\beta i} \tilde{K}_{\beta j} \quad \implies \quad c_{ij} = \frac{1}{M}\sum_{\beta = 1}^{K_{\text{tok}}} V_{\beta i} \tilde{K}_{\beta j}
\]
Finally, during the reading phase ($r = 1$) we select an index $\mu$ in the token sequence to run the neuron-astrocyte dynamics forward on. In other words, $c_{ij}$ is fixed across all tokens, but $I_j$ and $p_{ij}(0)$ change from token to token. For a particular index $\mu$ we instantiate the neurons $I_j$ \eqref{eq:I_eqs} and the astrocyte processes $p_{ij}$ as follows
\begin{equation}\label{eq:neuron_astro_trans_I_and_P}
\qquad I_j = \tilde{q}_j \equiv \tilde{Q}_{\mu j} \qquad \text{and} \qquad p_{ij}(0) = \tilde{Q}_{\mu j} \sum_{\sigma = 1}^{K_{tok}} \tilde{K}_{\sigma j}  \
\end{equation}
Plugging \eqref{eq:neuron_astro_trans_C} and \eqref{eq:neuron_astro_trans_I_and_P} into the neural fixed point condition for the reading phase \eqref{eq:x_trans_fp}, we arrive at the desired result
\[x^*_i = \frac{NM \sum\limits_{j=1}^M c_{ij} I_j}{\sum\limits_{i=1}^N \sum\limits_{j=1}^M p_{ij}(0)} = \frac{\frac{NM}{M}  \sum\limits_{j=1}^M \sum\limits_{\beta=1}^{K_\text{tok}} V_{\beta i} \tilde{K}_{\beta j} \tilde{Q}_{\mu j}}{\sum\limits_{i=1}^N \sum\limits_{j=1}^M \tilde{Q}_{\mu j} \sum\limits_{\sigma=1}^{K_\text{tok}} \tilde{K}_{\sigma j}} = \frac{\frac{NM}{M} \sum\limits_{j=1}^M \left(\sum\limits_{\beta=1}^{K_\text{tok}} V_{\beta i} \tilde{K}_{\beta j}\right) \tilde{Q}_{\mu j}}{N \sum\limits_{j=1}^M \tilde{Q}_{\mu j} \sum\limits_{\sigma=1}^{K_\text{tok}} \tilde{K}_{\sigma j}} \approx A_{\mu i}\]
which shows that for a particular choice of parameters and initialization, the neuron-astrocyte network converges to the output of self-attention. In other words, the neural fixed point equation \eqref{eq:x_trans_fp} is equal to the self-attention approximation \eqref{eq:approx_self_attn}.

\section{Details of Energy Network Experiments}\label{section:astro_energy_based}
\begin{figure}[ht!]
  \centering
  \includegraphics[width=\textwidth]{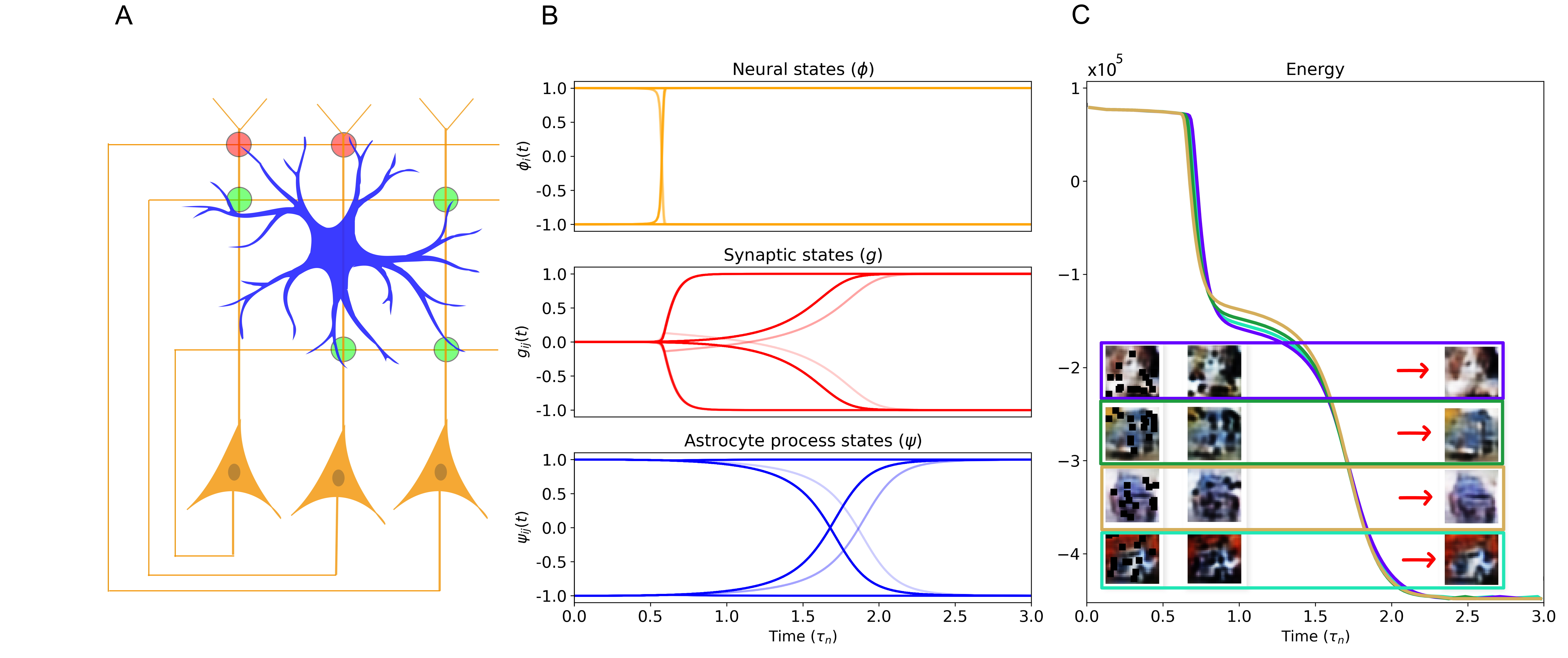}
  \caption{A) A schematic for our associative neuron-synapse-astrocyte network. B) The neural, synaptic, and astrocyte process activations during memory retrieval. In this case, the memory item being retrieved is an image of a dog taken from the CIFAR10 dataset. C) Decreasing energy function of the neuron-synapse-astrocyte network as the dynamics evolve. The decreasing energy functions during four different retrievals are shown.}
  \label{fig:cifar}
\end{figure}

To reduce the dimensionality of the problem, we use a custom autoencoder to encode the 3072-dimensional ($32 \times 32 \times 3$) CIFAR10 images into a smaller, 768 dimensional, latent space. A single CIFAR10 image in this latent space corresponds to a single memory $\xi^\mu$. In addition to being 768-dimensional, this latent space was also binary, so that $\xi^\mu \in [-1,1]^{768}$. To ensure that the latent space was binary, we wrote a custom autograd function which outputs the sign of the argument during the forward pass, but is linear during the backwards pass. The discrepancy between forward and backward pass induces a small amount of gradient noise in the training process, which is not significant enough to impair learning. For concreteness, in PyTorch this custom activation is given by:

\begin{lstlisting}[language=Python]
class RoundWithGradient(torch.autograd.Function):
    @staticmethod
    def forward(ctx, x):
        return torch.sign(x)

    @staticmethod
    def backward(ctx, grad_output):
        return grad_output


def round_with_gradient(x):
    return RoundWithGradient.apply(x)
\end{lstlisting}
To initialize the network, we reasoned (in analogy with traditional Hopfield networks) that the entire system should be initialized close to a stored memory. In our case, this includes all dynamical variables: neuron, synapses, and astrocytes. To do this, we set the time derivatives in \eqref{eq:neuron-astro-eqs} equal to zero, clamped the neural state at the corrupted memory $x_0$, and then solved the resulting set of algebraic equations for $p_{ij}(0)$ and $s_{ij}(0)$. 
Note that the synaptic states and process states are uniquely determined given a fixed neural state, due to the invertibility of $g$ and $\psi$.

\section{Details of Backpropagation Experiment}\label{section:astro_backprop}
To reduce the dimensionality of the problem, we assume that the the state of the processes does not depend on index $i$, in other words $p_{ij} = p_j$. Biologically, this has the interpretation that the astrocyte processes associated with post-synaptic neuron $i$ are all synchronized. This can be justified by assuming that nearby astrocyte processes are sensitive to inputs arrive at the dendritic tree of neuron $i$, and can rapidly redistribute their \ca levels. Similarly, we assume that the weights $T_{ijkl}$ between astrocyte processes $ij$ and $kl$ is only a function of indices $j$ and $l$. We likewise assume that the synapses only receive pre-synaptic input. That is,
\begin{align*}
\tau \ \dot{x}_i = -x_i + \sum_{j = 1}^N g_{ij} \phi_j + b_i \\ 
\tau \ \dot{s}_{ij} = -s_{ij} + \phi_j +  \psi_j \\
\tau \ \dot{p}_{j} = -p_{j} + \sum_{l = 1}^N T_{jl} \psi_l  + s_j \
\end{align*}
where $g_{ij} = W_{ij} \ \text{tanh}(s_{ij})$, $W_{ij}$ is a trainable parameter, and $\psi$ and $\phi$ are both also hyperbolic tangent. To match the dimensionality of the Tiny ImageNet dataset, our network contains $N =12288 = 64 \times 64 \times 3$ neurons. We numerically integrate the network using Euler integration for 20 timesteps, using a step-size of $\text{dt} = 0.1\tau$.  We set $\tau = 1$ in our experiment. As described in the main text, we initialized the neurons in the network as the masked images. The synapses and astrocyte processes we initialized at zero. The output of the network was a linear layer followed by a sigmoid function, to ensure valid RGB values. The network was trained using the Adam optimizer with a learning rate of $0.001$, using a batch size of 64 images. We trained on a subset of 5000 images in the TinyImage dataset, which enabled our network to learn quickly.


\end{document}